\documentclass{moriondNOTITLES}

\def\be{\begin{equation}}
\def\ee{\end{equation}}
\def\bea{\begin{eqnarray}}
\def\eea{\end{eqnarray}}

\newcommand{\Photo}{}

\usepackage[symbol]{footmisc}

\begin{document}
\vspace*{4cm}
\title{BSM reach of rare charm decays, including the rising star $\Lambda_c\to p \mu^+\mu^-$}

\author{Hector~Gisbert${}^{a,b,c}$, Gudrun~Hiller${}^{d}$, \underline{Dominik~Suelmann}${}^{d,}$\footnote[1]{Speaker}}

\address{${}^{a}$Escuela de Ciencias, Ingenier\'ia y Dise\~{n}o, Universidad Europea de Valencia, 46008 Valencia, Spain\\
${}^{b}$Istituto Nazionale di Fisica Nucleare (INFN), Sezione di Padova, 35131 Padova, Italy\\
${}^{c}$Dipartimento di Fisica e Astronomia `G.~Galilei', Universit\`a di Padova, 35131 Padova, Italy\\
${}^{d}$Fakult\"at f\"ur Physik, TU Dortmund, D-44221 Dortmund, Germany
}

\maketitle\abstracts{
 We perform the first global fit of rare charm transitions using 
 recent data on $D^0\to \mu^+\mu^-$, $D^+\to\pi^+\mu^+\mu^-$, $\Lambda_c\to p\mu^+\mu^-$ and 
 $D^0\to\pi^+\pi^-\mu^+\mu^-$ decays, including angular observables. We 
 work out constraints on new physics in the framework of the weak effective field theory. 
 While angular observables in $D^0\to\pi^+\pi^-\mu^+\mu^-$ decays provide sensitivities to different QCD models, 
 future null tests in $\Lambda_c\to p \mu^+\mu^-$ look more promising to extract limits because of 
 less hadronic uncertainties. This marks $\Lambda_c\to p \mu^+\mu^-$ as the rising star of rare charm decays.  
}

\section{Introduction}
Rare charm decays are notoriously challenging due to resonance pollution and poor 
convergence of effective field theories for charm decays. 
Not all is lost, however, and flavour-changing neutral currents (FCNCs) can still be 
utilized to cleanly signal new physics (NP) with the help of some special features of charm. 
FCNCs are suppressed in the Standard Model (SM) by the Glashow-Iliopoulos-Maiani (GIM) mechanism, which is 
especially efficient for up-type quarks. Notably, clean null test observables are available from angular distributions 
of $D$-mesons or $\Lambda_c$-baryons. 
We present the first global fit of rare FCNC charm 
decays with recent data on $D^0\to \mu^+\mu^-$, $D^+\to\pi^+\mu^+\mu^-$, $\Lambda_c\to p\mu^+\mu^-$ and 
$D^0\to\pi^+\pi^-\mu^+\mu^-$, including angular observables with GIM-protected null tests. 
While the null tests are able to cleanly signal NP, the interpretation of the data in terms 
of extracting limits on the Wilson coefficients suffers from hadronic uncertainties including 
unknown strong phases.
\section{Model Setup}
We use the weak effective theory framework for  semileptonic $|\Delta c| = |\Delta u| = 1$ transitions. 
They  are described  by the  effective  low energy Hamiltonian
\begin{equation}
       \mathcal{H}_{\rm eff} \subset -\frac{4\,G_F}{\sqrt{2}} \frac{\alpha_e}{4\pi} \sum_i \bigg(\mathcal{C}_i(\mu) \mathcal{O}_i + \mathcal{C}_i(\mu)^\prime \mathcal{O}_i^\prime  \bigg) \:,
\end{equation}
\vspace{-0.2cm}
\begin{equation} \label{eq:operators}
       \renewcommand{\arraystretch}{1.2}
       \begin{array}{r@{\,}lr@{\,}lr@{\,}l}
              \mathcal{O}_1^{(q)} &= (\bar{u}_L \gamma_\mu T^a q_L)(\bar{q}_L \gamma^\mu T^a c_L)~,  &
              \mathcal{O}_2^{(q)} &= (\bar{u}_L \gamma_\mu q_L)(\bar{q}_L \gamma^\mu c_L)~, &  &\\
              \mathcal{O}_7 &= \frac{m_c}{e} (\bar{u}_L \sigma_{\mu\nu} c_R)F^{\mu\nu}~,  &
              \mathcal{O}_9 &= (\bar{u}_L \gamma_\mu c_L)(\bar{\ell} \gamma^\mu \ell)~,  &
              \mathcal{O}_{10} &= (\bar{u}_L \gamma_\mu c_L)(\bar{\ell} \gamma^\mu \gamma_5 \ell)~. 
       \end{array}
\end{equation}
Here, $q_{L,R} = \frac{1}{2}(1 \mp \gamma_5)q$ represent the chiral quark fields. 
Primed operators $\mathcal{O}_i^\prime$ are obtained from the $\mathcal{O}_i$ by swapping the chiral projectors of the quark fields $L(R)\to R(L)$.
In the SM the coefficients $\mathcal{O}_{1,2}^{(q)}$ are order one, and others suppressed, with primed coefficients 
strongly suppressed by $m_u/m_c$. The special feature of charm is here that also the coefficient of 
$\mathcal{O}_{10}$ remains GIM protected, $\mathcal{C}_{10}^{\mathrm{SM}} = 0$.
The operators $\mathcal{O}_{7,9}$ receive perturbative contributions mainly from $\mathcal{O}_{1,2}^{(q)}$ included in effective coefficients,
which are however small and subdominant for SM phenomenology.
To model the challenging resonance contributions for both $D^0\to R_1(\to\pi^+\pi^-)R(\to \mu^+\mu^-)$ and $\Lambda_c \to p R(\to \mu^+\mu^-)$, 
$R_1=\rho/\omega, f_0(500)$, $R=\rho,\omega,\phi$, we employ a phenomenological ansatz
\begin{equation}
       \mathcal{C}_9^{\mathcal{R}}(q^2) = \frac{a_\rho \mathrm{e}^{i\delta_\rho}}{q^2-m_\rho^2 + i m_\rho \Gamma_\rho} + \frac{a_\omega \mathrm{e}^{i\delta_\omega}}{q^2-m_\omega^2 + i m_\omega \Gamma_\omega} + \frac{a_\phi \mathrm{e}^{i\delta_\phi}}{q^2-m_\phi^2 + i m_\phi \Gamma_\phi}
       \label{eq:C9R}
\end{equation}
with the resonance masses $m_{R}$, decay widths $\Gamma_R$ and 
the a priori unknown process-dependent strong phases $\delta_R$ and resonance parameters 
$a_R \geq 0$. 
For the individual resonance contributions to $D^0\to\pi^+\pi^-\mu^+\mu^-$ we use separate resonance parameters, 
increasing the number of resonance parameters to $12$ for this mode.
\section{Four-body decay \texorpdfstring{$D^0\to \pi^+\pi^-\mu^+\mu^-$}{D0 -> pi pi mu mu}}
The 5-differential decay distribution of $D^0\to\pi^+\pi^-\mu^+\mu^-$ can be written as~\cite{DeBoer:2018pdx}
\begin{equation}   \label{eq:full}
              \frac{\mathrm{d}^5\Gamma}{\mathrm{d}q^2\,\mathrm{d}p^2\,\mathrm{d}\cos\theta_{P_1}\,\mathrm{d}\cos\theta_\ell\, \mathrm{d}\phi} 
       =\frac{1}{ 2 \, \pi}  \sum_{i=1}^9 c_i(\theta_\ell,\phi)\, I_i (q^2,p^2,\cos \theta_{P_1})\,,
\end{equation}
where $q^2$ and $p^2$ represent the invariant mass-squared of the dilepton and the ($\pi^+ \pi^-$)-subsystem, respectively. 
$\theta_\ell$, $\theta_{P_1}$, 
and $\phi$ are the angular variables~\footnote[7]{Our convention differs from the experimental analysis~\cite{LHCb:2021yxk} 
and $\phi=\pi+\phi_\mathrm{LHCb}$. 
Resulting in $I_{4,5,7,8}=- I_{4,5,7,8}^\mathrm{LHCb}$. \label{foot:angles} } defined in ref.~\cite{DeBoer:2018pdx}. 
The angular coefficients $I_i$ depend on $q^2$, $p^2$, and $\theta_{P_1}$, 
while the dependence on the other angles is encoded in the $c_i$ functions. 
Following the notation in ref.~\cite{LHCb:2021yxk} we construct CP-symmetries $\langle S_i \rangle$ 
and asymmetries $\langle A_i \rangle$ related to the angular coefficients $I_i$. % via 
Finite $I_{5,6,7}$ and finite CP-asymmetries require non-vanishing axial-vector contributions 
and CP-violating phases respectively. These are however very small in the SM because 
of the GIM mechanism and the involved CKM phases, hence $\langle S_{5,6,7} \rangle$ 
and all CP-asymmetries serve as clean null tests of the SM~\cite{DeBoer:2018pdx}. 

\begin{figure}\centering
       \begin{minipage}{0.45\linewidth}
       \centerline{\includegraphics[width=0.75\linewidth]{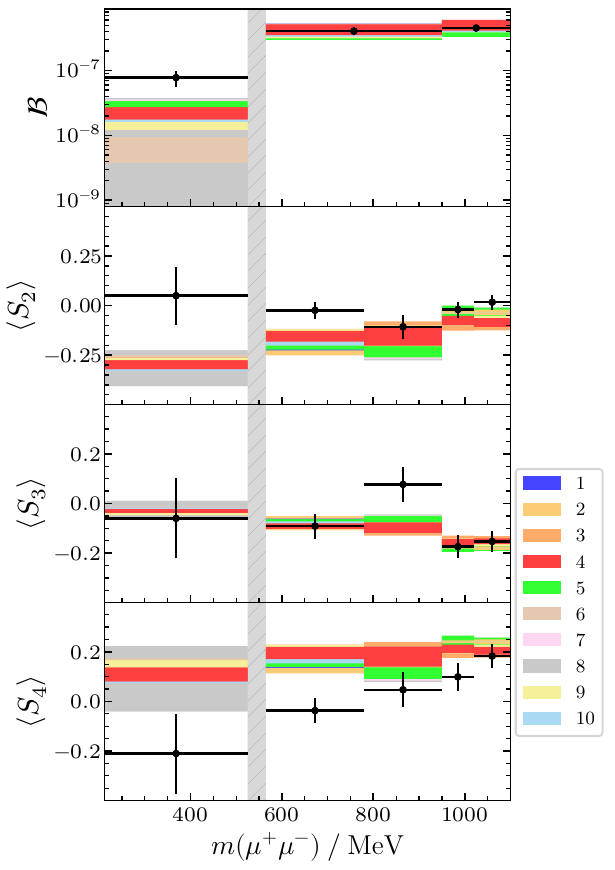}}
              \caption[]{
                     %Data (black) on the $D^0\to\pi^+\pi^-\mu^+\mu^-$ binned branching ratio~\cite{LHCb:2017uns} 
                     %and angular observables~\cite{LHCb:2021yxk} $\langle S_{2,3,4} \rangle$        
              Data~\cite{LHCb:2017uns,LHCb:2021yxk} (black) of $\mathcal{B}$ and angular observables of $D^0\to\pi^+\pi^-\mu^+\mu^-$ decays
              and SM predictions for different fit scenarios. Figure from \cite{Gisbert:2024kob}.}
              \label{fig:S234}
       \end{minipage}
       \hfill
       \begin{minipage}{0.53\linewidth}
              \centerline{
                     \includegraphics[width=0.49233\linewidth]{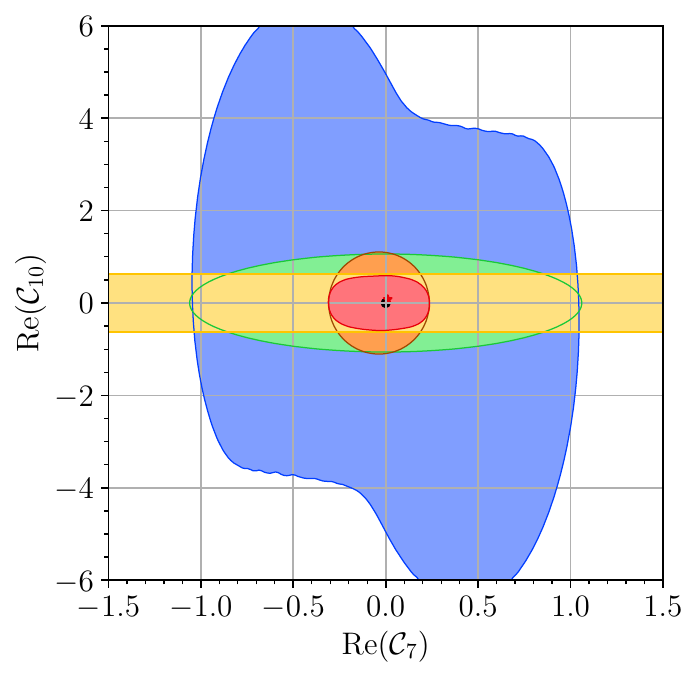} %0.422
                     }
              \centerline{
                     \includegraphics[width=0.7\linewidth]{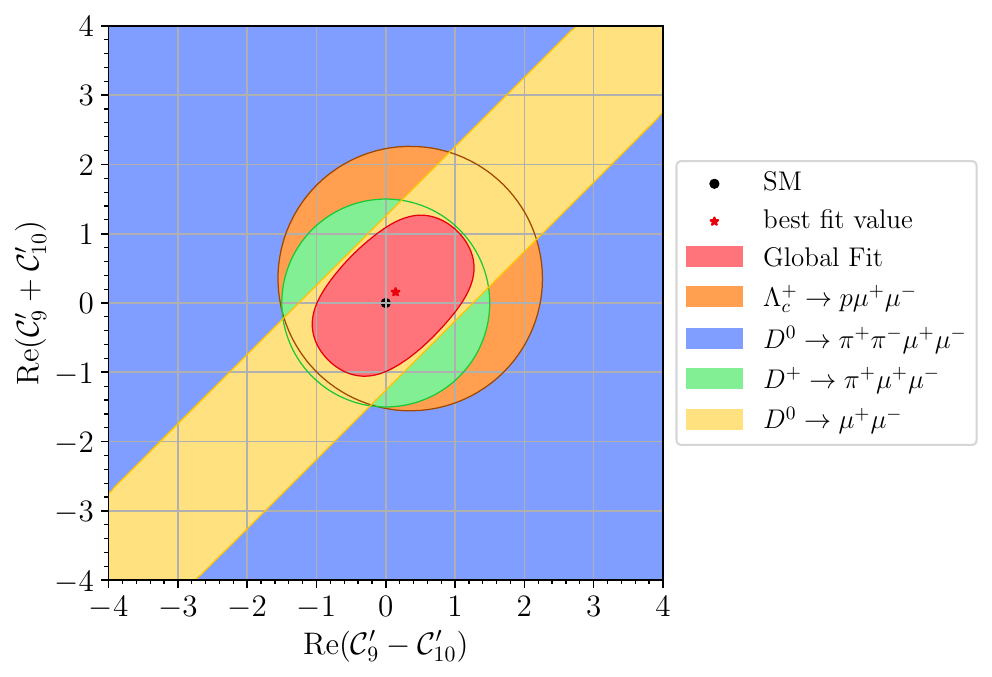} %0.6
                     }
                     \caption[]{Contours of the 2D fits to $(\mathrm{Re}\mathcal{C}_{7},\mathrm{Re}\mathcal{C}_{10})$ 
and $(\mathrm{Re}(\mathcal{C}_{9}^\prime - \mathcal{C}_{10}^\prime), \mathrm{Re}(\mathcal{C}_{9}^\prime + \mathcal{C}_{10}^\prime))$. 
Figures from \cite{Gisbert:2024kob}. }
              \label{fig:contours}
       \end{minipage}
\end{figure}

We determine the resonance parameters in the ansatz of eq.~\ref{eq:C9R} from the available measurements in ref.~\cite{LHCb:2017uns,LHCb:2021yxk}, 
including branching ratios, angular symmetries $\langle S_{2,3,4} \rangle$ and differential distributions $\mathrm{d}\Gamma/\mathrm{d}\sqrt{q^2}$, 
$\mathrm{d}\Gamma/\mathrm{d}\sqrt{p^2}$. The latter two are however unfortunately provided without systematic uncertainties, accounting for unfolded detector effects.
We perform fits in scenarios with different resonance models and data sets, detailed in ref.~\cite{Gisbert:2024kob}.
The CP-averaged angular observables $\langle S_{2,3,4} \rangle$ together with the binned branching ratio using the fit value 
of the resonance parameters for the different scenarios are displayed in fig.~\ref{fig:S234}. 
We observe some tensions with the data for the first two low-$q^2$ bins, consistent with~\cite{Fajfer:2023tkp}, 
likely caused by additional contributions beyond our resonance ansatz and beyond the scope of our work. For NP fits we exclude those bins.
To improve on the resonance model, to separate the different resonances and 
to include potential additional ones in the future, 
we suggest to provide the double differential branching ratio $\mathrm{d}^2\mathcal{B}/\mathrm{d}q^2\mathrm{d}p^2$ including bin-to-bin correlations in a future measurement. 

We perform global fits to NP driven by $\mathcal{C}_{7,9,10}$ using the best-fit values of 
hadronic parameters for $D^0\to\pi^+\pi^-\mu^+\mu^-$ for scenario 4 (red), shown in fig.~\ref{fig:S234}. 
Our results~\cite{Gisbert:2024kob} indicate that bounds on $\mathcal{C}_{7,9}$ are somewhat dependent 
on the fit scenario of the resonance parameters, while the bounds on $\mathcal{C}_{10}$ are overall stable.
For the NP constraint from $D^0\to\pi^+\pi^-\mu^+\mu^-$ we include both the branching ratios for normalization 
and all available null tests. We further include additional constraints from upper limits on the branching ratio $\mathcal{B}$
for $D^0\to\mu^+\mu^-$ and low-$q^2$ and high-$q^2$ upper limits for $D^+\to\pi^+\mu^+\mu^-$ and $\Lambda_c \to p \mu^+\mu^-$.
In fig.~\ref{fig:contours} we show the allowed regions for 2D fits of $(\mathrm{Re}\mathcal{C}_{7},\mathrm{Re}\mathcal{C}_{10})$ 
and $(\mathrm{Re}(\mathcal{C}_{9}^\prime - \mathcal{C}_{10}^\prime), \mathrm{Re}(\mathcal{C}_{9}^\prime + \mathcal{C}_{10}^\prime))$.
For further fit results see~\cite{Gisbert:2024kob}. The global fit is dominated by $2$- and $3$-body 
decays and $D^0\to\pi^+\pi^-\mu^+\mu^-$ is not competitive, because of the added complexity of a $4$-body decay, 
including additional strong phase uncertainties 
and the current experimental sensitivity of the null tests.
\section{Rising star \texorpdfstring{$\Lambda_c \to p \mu^+\mu^-$}{Lc -> p mu mu}}
\begin{figure}
       \begin{minipage}{0.46\linewidth}
       \centerline{\includegraphics[width=\linewidth]{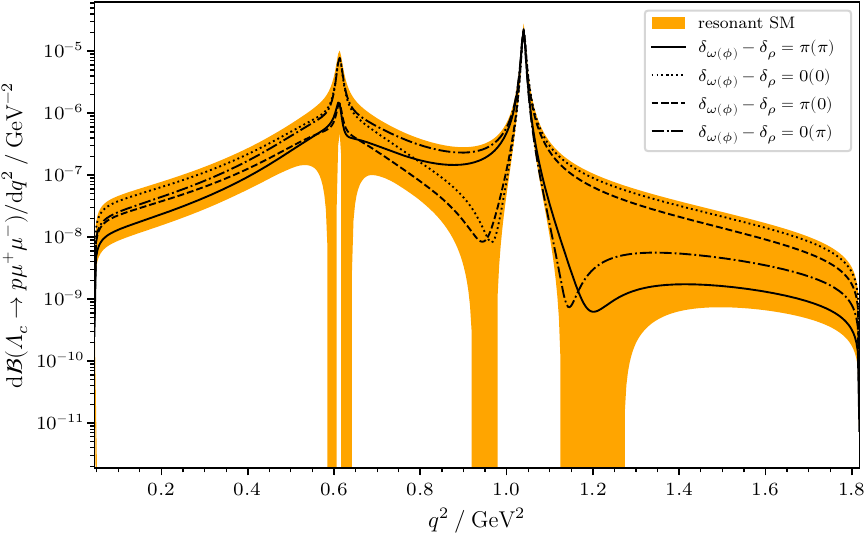}}
       \caption[]{The differential branching ratio of $\Lambda_c\to p \mu^+\mu^-$ decays including the uncertainties from strong phases (orange band)
       and curves for fixed phases (black). Figure from \cite{Gisbert:2024kob}.}
       \label{fig:dBR_Lambdac}
       \end{minipage}
       \hfill
       \begin{minipage}{0.46\linewidth}
       \centerline{\includegraphics[width=\linewidth]{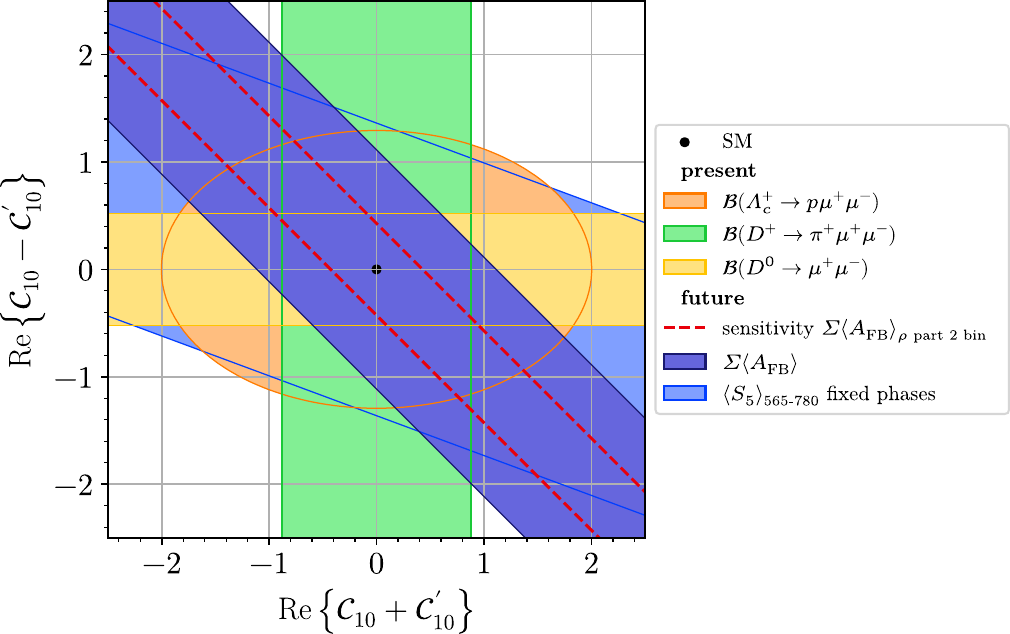}}
       \caption[]{Constraints of $\mathrm{Re}(\mathcal{C}_{10}\pm\mathcal{C}_{10}^\prime)$ from 
       current branching ratio data together with hypothetical future measurements of null tests. Figure from \cite{Gisbert:2024kob}.}
       \label{fig:C10complementarity}
       \end{minipage}
\end{figure}

The double differential distribution for $\Lambda_c \to p \ell^+\ell^-$ can be written as~\cite{Golz:2021imq}
\begin{equation}
       \frac{\mathrm{d}^2\Gamma}{\mathrm{d}q^2\mathrm{d}\cos\theta_\ell} 
       = \frac{3}{2} \left(K_{1ss}(q^2)\sin^2\theta_\ell + K_{1cc}(q^2)\cos^2\theta_\ell + K_{1c}(q^2)\cos\theta_\ell\right)
\end{equation}
The angular functions $K_{1ss}$, $K_{1cc}$ and $K_{1c}$ include contributions from various NP coefficients, see ref.~\cite{Golz:2021imq}.
The baryonic decay mode is theoretically simpler than $4$-body decays due to fewer strong phases and available lattice form factors~\cite{Meinel:2017ggx}, 
and still phenomenological richer than $D\to P\ell^+\ell^-$ due to the higher spin.
Importantly, in $\Lambda_c\to p \mu^+\mu^-$ the forward-backward asymmetry 
\begin{equation}
       \langle A_{\mathrm{FB}} \rangle = \frac{1}{\Gamma} \int_{q^2_{\mathrm{min}}}^{q^2_{\mathrm{max}}} \left(\int_0^1-\int_{-1}^0\right) \frac{\mathrm{d}^2\Gamma}{\mathrm{d}q^2\,\mathrm{d}\!\cos\theta_\ell} \mathrm{d}q^2\,\mathrm{d}\!\cos\theta_\ell = \frac{3}{2} \frac{1}{\Gamma} \int_{q^2_{\mathrm{min}}}^{q^2_{\mathrm{max}}}\,K_{1c}(q^2)\,\mathrm{d}q^2
\end{equation}
with $A_{\mathrm{FB}} \propto K_{1c} \propto \mathcal{C}_{10}$ 
serves as a clean null test of the SM.
The resonance parameters $a_R$ in the ansatz of eq.~\ref{eq:C9R} 
are determined using branching ratio data in the $q^2$-regions of ref.~\cite{LHCb:2024hju}, 
while a fit of the relative strong phases $\delta_{\phi(\omega)} - \delta_\rho$ requires finer 
binning in regions, where resonances overlap. In fig.~\ref{fig:dBR_Lambdac} we show the differential branching ratio, including
curves (black) with fixed relative strong phases. 
After our publication~\cite{Gisbert:2024kob} additional data has been made available in ref.~\cite{LHCb:2024hju,LHCb:2025bfy}. 
Background subtracted dimuon mass spectra have been provided allowing one to estimate first constraints 
on the relative strong phases. 

To estimate the future sensitivity of $\Lambda_c \to p\ell^+\ell^-$ we consider 
the CP-asymmetry $\Delta\langle A_{\mathrm{FB}}\rangle$ and CP-symmetry $\Sigma\langle A_{\mathrm{FB}}\rangle$ of 
the forward-backward asymmetry. 
For NP both are enhanced by the interference between resonant SM contributions and NP.
Therefor, similar to $D^0\to\pi^+\pi^-\mu^+\mu^-$, they depend 
strongly on the strong phases. 
Because of the normalization of $\langle A_\mathrm{FB}\rangle$ to the branching ratio,
generically the asymmetry is larger for bins with a smaller branching ratio, and vice versa.
We explore the $q^2$ binning to find those with the strongest enhancement and different dependencies on the strong phases. 
We suggest "optimized" bins that have a branching 
ratio at a reasonable level and have the highest sensitivity to NP. 
In fig.~\ref{fig:C10complementarity} we show current constraints from branching ratios as well as for future scenarios, detailed in ref.~\cite{Gisbert:2024kob}. 
For $\Sigma\langle A_\mathrm{FB}\rangle$ we also indicate the maximal sensitivity to NP (red dashed), which 
a measurement could reach. Even for values of the strong 
phases that give the maximal enhancement, the values of the null tests will be below the sensitivity of our scenario.  
As apparent from fig.~\ref{fig:C10complementarity}, the global fit profits from synergies between 
the various decays and observables, with each observable providing constraints in different directions.
Our future scenario for $A_{\mathrm{FB}}$ assumes a similar precision 
as its recent measurement~\cite{LHCb:2025bfy} with however a different $q^2$-binning. This indicates that these null tests 
are within experimental reach and can contribute to the global fit via complementary directions.  

\section{Conclusion}

We perform the first global fit of rare charm decays using branching 
ratio data of various rare charm 
decays, and for the first time, angular observables, including the null tests of $D^0\to\pi^+\pi^-\mu^+\mu^-$. 
The many GIM-protected null test observables of this decay 
are strongly enhanced by the interference between NP and resonant SM amplitudes. 
Despite this, because 
of its complicated hadronic structure, $D\to \pi^+\pi^-\mu^+\mu^-$ does not currently play a role in the global fit. It has 
however a good potential to probe QCD frameworks and hadronic dynamics.
The decay $\Lambda_c\to p\mu^+\mu^-$ on the other hand is the rising star with the simplicity of a 3-body 
decay, fewer hadronic parameters, sensitivity to various NP operators and featuring a null 
test distribution. First measurements of $A_{\mathrm{FB}}$ have even been performed at relevant precision ~\cite{LHCb:2025bfy}.

\section*{Acknowledgments}

DS would like to thank the organizers of the $59^{\mathrm{\tiny th}}$  Rencontres de Moriond EW 2025. 
This work is supported by the \textit{Bundesminsterium für Bildung und Forschung} (BMBF).

\section*{References}
\bibliography{moriond}

\end{document}